\documentstyle[epsf,eqsecnum,floats,preprint,aps]{revtex}
\catcode`@=11
\def\gtrsim{\mathrel{\mathpalette\vereq>}}
\def\lesssim{\mathrel{\mathpalette\vereq<}}
\def\vereq#1#2{\lower3pt\vbox{\baselineskip1.5pt \lineskip1.5pt
\ialign{$\m@th#1\hfill##\hfil$\crcr#2\crcr\sim\crcr}}}
\catcode`@=12 
\tighten

\begin{document}
\rightline{CU-TP-1019}
\rightline{gr-qc/0106030}
\vskip 1cm

\begin{center}
\ \\
\large{{\bf Black holes with hair}}\footnote{Lectures given at the
International School Of Cosmology and Gravitation: 17th Course:
Advances in the Interplay between Quantum and Gravity Physics,  Erice,
Italy, May 2001.}

\ \\
\ \\
\normalsize{Erick J. Weinberg\footnote{\tt ejw@phys.columbia.edu}}
\ \\
\small{\em Department of Physics \\
Columbia University \\
New York, NY 10027}

\end{center}

\setcounter{page}{1}
\thispagestyle{empty}
\baselineskip 16pt plus 2pt minus 2pt

\section{Introduction}

Among the most remarkable results of classical general relativity are
the black hole uniqueness theorems for pure gravity and for gravity
coupled to electromagnetism.  The simplicity and elegance of these
black holes inspired Chandrasekhar's statement, in the prologue to
his treatise \cite{Chandra}, that ``the black holes of nature are the
most perfect macroscopic objects there are in the universe \dots and
\dots they are the simplest objects as well''.  

These uniqueness theorems, together with related results on  
black holes coupled to other types of matter and on the
behavior of matter as it collapses to form a black hole, led to the
widely repeated statement that ``black holes have no hair''
\cite{MTW}.  This statement had various interpretations.  Some took it
to mean that the only possible static fields outside a black hole
horizon are those required by the conserved long-range charges.  A
weaker interpretation allowed such ``hair'', but required that the
solution be uniquely determined by its mass, angular momentum, and
conserved charges.  In either case, there was a question of whether
the statement applied to all solutions, or only to stable solutions.

Many in the wider theoretical physics community thought that a general
result, restricted perhaps by technical assumptions, had been established.
In fact, as was clear to experts in the field, no-hair theorems had
only been proven for very specific types of matter, and the more
general statement, however interpreted, was only a conjecture.

Over the last decade it has become clear that this conjecture, even in
its weaker form, is not in general true.  When gravity is coupled
to matter theories that have more complex structures --- including
theories similar to those of the standard model --- there are black
hole solutions that do, indeed, have hair.  These black holes are most
naturally subatomic, rather than astrophysical, in size.  Interesting
in their own right, they also help clarify which features are general
characteristics of classical black holes and which are not, and at the
same time lend insight into the quantum mechanical connection
between black hole dynamics and thermodynamics.

By now, a variety of solutions with hair are known.  In these lectures
I will focus on the magnetically charged black holes that arise in
spontaneously broken Yang-Mills-Higgs theories and on
the properties of the related self-gravitating nonsingular magnetic
monopoles.  For an extensive review that includes discussions of other
types of solutions, see \cite{Volkov:1999cc}.  

After a brief general discussion of spherically symmetric black holes, I
will review some of the properties of 't Hooft-Polyakov monopoles in flat
spacetime.  While these are usually understood from a topological point of
view, I will present some energetic arguments that are perhaps more helpful
in understanding the related black hole solutions.  I will then describe the
effects of gravity on the singly-charged monopole.  These are two-fold. 
First, there is an upper bound on the mass of a nonsingular monopole, with
the monopole going over into an extremal black hole as this limit is
reached.  Second, it is possible to embed a black hole within
the monopole core, thus yielding a black hole with hair.  These 
solutions with hair can be degenerate in mass and charge with pure 
Reissner-Nordstrom solutions.  I will show that in these theories 
the latter have a classical instability that leads to decay into solutions 
with hair.  In the case of Reissner-Nordstrom black holes with higher 
magnetic charge, this instability can lead to static black holes without 
any rotational symmetry.  Finally, in the last part of these lectures I 
will examine in more detail the transition from nonsingular monopole 
to black hole, focusing on how the ``quasi-black holes'' that are just 
short of this transition appear to an external observer.  As we will see, 
these provide interesting insights into the origin of black hole entropy.

\section{Spherically symmetric black holes}

For the sake of simplicity, in these lectures I will focus for the
most part on solutions with static, spherically symmetric metrics.
Any such metric can be written in the form
\begin{equation}
    ds^2 = - B(r) dt^2 + A(r) dr^2 + R^2(r) (d\theta^2 + \sin^2\theta
    d\phi^2)   \, .
\end{equation}
Furthermore, I will use the freedom to redefine coordinates to set 
$R(r) = r$ and write 
\begin{equation}
    ds^2 = - B(r) dt^2 + A(r) dr^2 + r^2 (d\theta^2 + \sin^2\theta
    d\phi^2)   \, .
\label{sphsymmetric}
\end{equation}
A zero of $1/A$ corresponds to a horizon, while a double zero
corresponds to an extremal horizon.

The two simplest black holes of this form are the Schwarzschild
and the Reissner-Nordstrom solutions.  The Schwarzschild black hole is
a vacuum solution with
\begin{equation}
     B_{\rm Sch} = A_{\rm Sch}^{-1} = 1 -{2 MG \over r}   \, .
\end{equation}
There is a coordinate singularity at the horizon, $r = 2GM$, and a
true curvature singularity at $r=0$.  The maximally extended spacetime
contains two exterior regions, each with $2MG<r < \infty$ and
$-\infty< t < \infty$; a region, with $0< r<2MG$, that lies to the
future of the horizon and ends on a spacelike $r=0$ singularity; and
finally a region, also with $0< r<2MG$, that lies in the past of the
horizon and has an initial spacelike $r=0$ singularity.  It is
important to keep in mind that $r$ is actually a timelike coordinate
for values less than $2MG$.  Hence, it is somewhat misleading to think
of the region with $r < 2MG$ as the ``interior'' of the black hole;
one can draw a complete spacelike hypersurface through the spacetime
on which $r$ is never less than $2MG$.

The Reissner-Nordstrom solution has Coulomb electric and magnetic
fields
\begin{equation}
     {\bf E} = Q_{\rm E} \,{ {\bf \hat r}\over r^2}   \qquad \qquad
     {\bf B} = Q_{\rm M} \, {{\bf \hat r}\over r^2} 
\end{equation}
and a metric
\begin{equation}
     B_{\rm RN} = A_{\rm RN}^{-1} 
   = 1 -{2 MG \over r} 
      + {4\pi G ({Q_{\rm E}^2 + Q_{\rm M}^2}) \over r^2}  \, .
\end{equation}
There are three cases to consider.  If 
\begin{equation}
     M > \sqrt{4\pi (Q_{\rm E}^2 + Q_{\rm M}^2)}\, M_{\rm Pl}
\end{equation}
(where the Planck mass $M_{\rm Pl} = G^{-1/2}$ in units where 
$c=\hbar =1$) the metric describes a black hole solution
with horizons at 
\begin{equation}
      r_\pm = MG \pm \sqrt{M^2G^2 - 4 \pi G (Q_{\rm E}^2 + Q_{\rm M}^2)}
\end{equation}
and a timelike curvature singularity at $r=0$.  The maximally extended
spacetime contains an infinite sequence of exterior regions.  It is
possible for a worldline to pass through an infinite sequence of such
regions without ever encountering the $r=0$ singularities.

If 
\begin{equation}
     M = \sqrt{4\pi (Q_{\rm E}^2 + Q_{\rm M}^2)} \, M_{\rm Pl}
\end{equation}
there is an extremal black hole, with a horizon at 
\begin{equation}
      r_0 = MG = \sqrt{4\pi (Q_{\rm E}^2 + Q_{\rm M}^2)} \, M_{\rm
      Pl}^{-1}   \, . 
\end{equation}
As in the previous case, the maximally extended spacetime contains an
infinite sequence of exterior regions, and it is possible to avoid the
timelike singularity at $r=0$.  On any hypersurface of constant time,
the extremal horizon at $r=r_0$ is an infinite proper distance from
any point with $r\ne r_0$; nevertheless, a worldline starting at any
$r> r_0$ can cross the horizon and reach $r< r_0$ in a finite
proper time.

Finally, if 
\begin{equation}
     M < \sqrt{4\pi (Q_{\rm E}^2 + Q_{\rm M}^2)} \, M_{\rm Pl}
\end{equation}
there is no horizon, but only a naked singularity at $r=0$.  

The Schwarzschild and Reissner-Nordstrom solutions are the only static
black hole solutions in the Einstein-Maxwell theory; if we only
require that the solution be stationary, there is also the Kerr-Newman
solution, which includes the others as special cases.  Thus, these
black holes are completely specified by giving their mass, angular
momentum, and electric and magnetic charges. This result was the
inspiration for the no-hair conjecture.  However, although the
statement that ``black holes have no hair'' was widely repeated, this
conjecture was actually proven only in a number of very specific
contexts.

As an example of these, consider the case of gravity coupled to a
scalar field $\phi(x)$ \cite{Bekenstein:1972hc}.  The dynamics of
$\phi$ are governed by a potential $V(\phi)$ that is assumed to have a
single minimum, at $\phi = \phi_0$.  To simplify the presentation I
will assume spherical symmetry, although the proof is readily extended
to the more general case.  With a metric of the form given in
Eq.~(\ref{sphsymmetric}), and $\phi$ assumed to depend only on $r$,
the scalar field equations take the form
\begin{equation}
    {1 \over r^2 \sqrt{AB}} \left({r^2 \sqrt{AB}\, \phi' \over A}
    \right)' = {dV \over d\phi}
\end{equation}
with primes denoting differentiation with respect to $r$.  Multiplying
both sides of this equation by common factors, we obtain
\begin{equation}
     (\phi - \phi_0) \left({r^2 \sqrt{AB} \,\phi' \over A} \right)' 
    = (\phi - \phi_0)\, r^2\,\sqrt{AB}\,\, {dV \over d\phi}  \, .
\end{equation}
We now assume that there is a horizon at $r=r_{\rm H}$, and integrate the
above equation from $r_{\rm H}$ to infinity.  An integration by parts leads
to 
\begin{equation}
    \int_{r_{\rm H}}^\infty dr \, 
   {d\over dr}\left[(\phi - \phi_0)\,{r^2 \sqrt{AB}\, \phi' \over A}
   \right]
     =  \int_{r_{\rm H}}^\infty dr \,r^2 \sqrt{AB}\left[ {(\phi')^2 \over A} 
        + (\phi - \phi_0)\,{dV \over d\phi} \right]   \, .
\end{equation}
The left hand side is equal to the sum of surface terms at the horizon
and at infinity.  The former vanishes because $1/A=0$ on the
horizon.  Because $\phi$ must approach its vacuum value at $r=\infty$, the
decreases in $\phi'$ and $\phi - \phi_0$ are rapid enough that the
surface term at infinity also vanishes. The integral on the right hand
side must therefore be equal to zero.  The first term in the integrand is
manifestly positive (since $A>0$ outside the horizon), while our
assumption that $V$ has a single minimum implies that the second term
is also positive.  Hence, the only way that the integral can vanish
is for $\phi(r)$ to be equal to its vacuum value $\phi_0$ everywhere
outside the horizon.  

This proof relied crucially on the assumed properties of $V(\phi)$.
It would have failed if the potential had multiple minima, or if there
were additional fields present.  Although the proof can be extended to
a wider class of scalar field theories \cite{Bekenstein:1995un}, this
reliance on the details of the theory suggests that it might be
possible to construct black holes with hair in a theory with a
sufficiently complex structure.  As we will see, a natural place to
look is the spontaneously broken gauge theories that support magnetic
monopole solutions in flat spacetime.

\section{Magnetic monopoles in flat spacetime}
\label{sec:monopole}

Consider an SU(2) Yang-Mills theory with a triplet scalar
field $\phi^a$ and a Lagrangian
\begin{equation}
    {\cal L} = -{1\over 4} (F_{\mu\nu}^a)^2 + {1\over 2} (D_\mu \phi)^2
       - V(\phi)
\end{equation}
where the field strength
\begin{equation}
    F_{\mu\nu}^a = \partial_\mu A_\nu^a - \partial_\nu A_\mu^a
      - e\,\epsilon_{abc} A_\mu^b A_\nu^c  \, ,
\end{equation}
the covariant derivative 
\begin{equation}
   D_\mu \phi^a = \partial_\mu \phi^a 
        - e \,\epsilon_{abc} A_\mu^b \phi^c  \, ,
\end{equation}
and the scalar field potential 
\begin{equation}
    V(\phi) = -{\mu^2 \over 2} \phi^2  + {\lambda\over 2} 
      (\phi^2)^2 
\end{equation}
with $\mu^2$ and $\lambda$ both positive.

The potential has a family of gauge-equivalent minima
with 
\begin{equation} 
    \phi^2  = v^2 \equiv {\mu^2 \over 2\lambda}
\end{equation}
that spontaneously break the SU(2) symmetry down to U(1). Without loss
of generality, we can choose the vacuum with $\phi^a = v \delta^{a3}$.
The fields corresponding to the physical elementary particles are then
the ``electromagnetic'' U(1) gauge field ${\cal A}_\mu = A_\mu^3$, a
neutral scalar field $\varphi = \phi^3$, and a complex vector field
$W_\mu = (A_\mu^1 + i A_\mu^2)/\sqrt{2}$ whose quanta are spin-one
particles with electric charge $\pm e$ and mass $m_W=ev$.  In terms of
these fields, the Lagrangian can be written as
\begin{eqnarray} 
    {\cal L} &=& -{1\over 2}\left|{\cal D}_\mu W_\nu 
         - {\cal D}_\nu W_\mu \right|^2 
       -{1\over 4}({\cal F}_{\mu\nu})^2
       +{1\over 2}d_{\mu\nu}{\cal F}^{\mu\nu}
     -{1\over 4}(d_{\mu\nu})^2  \cr && \qquad
      + {e^2\varphi^2}|W_\mu|^2
        + {1 \over 2}(\partial_\mu \varphi)^2 - V(\varphi)
\label{unitaryLag}
\end{eqnarray}
where 
\begin{equation} 
   {\cal F}_{\mu\nu} = \partial_\mu {\cal A}_\nu 
       - \partial_\nu {\cal A}_\mu
\end{equation}
and
\begin{equation} 
   {\cal D}_\mu  = (\partial_\mu - i e {\cal A}_\mu)
\end{equation}
denote the electromagnetic field strength and covariant derivative 
and 
\begin{equation} 
    d_{\mu\nu} = ie [W_\mu^*W_\nu - W_\nu^*W_\mu] 
\end{equation}
is the magnetic moment density due to the charged vector field. 

This theory possesses finite energy magnetic monopole solutions
\cite{'tHooft:1974qc,Polyakov:1974ek}.  Their
existence is usually motivated by topological arguments.  One begins by
considering configurations in which the scalar field at spatial infinity
has its SU(2) orientation correlated with the direction in space, so that
as $r \to \infty$
\begin{equation} 
      \phi^a \rightarrow  v \hat r^a    \, .
\label{phitopology}
\end{equation}
Because such a configuration cannot be smoothly deformed to the uniform
vacuum solution, it should be possible to obtain a static solution by
minimizing the energy subject to this boundary condition.  
In order that the energy be finite, $D_i\phi$ must fall faster than
$r^{-3/2}$, which implies a vector potential 
\begin{equation} 
    A_i^a =  \epsilon_{iak}\, {\hat r^k \over er} + O(r^{-2})
\end{equation}
that gives rise to a Coulomb magnetic field
\begin{equation} 
    B_i^a = {\hat r^a \hat r^k \over er^2}  + O(r^{-3})   \, .
\end{equation}
Thus, this configuration describes a magnetic monopole with magnetic
charge $Q_{\rm M} = 1/e$.  Higher charges can be obtained by allowing
additional twisting of the asymptotic scalar field, but these must
obey the topological quantization condition 
\begin{equation} 
     Q_{\rm M} = {n \over e}    \, .
\end{equation}

One can proceed further by adopting the Ansatz
\begin{eqnarray}
    \phi^a &=& v \,\hat r^a\, h(r)  \cr 
    A_i^a &=& \epsilon_{iak} \,\hat r^k \left[{1-u(r) \over er}\right] 
\label{monoAnsatz}
\end{eqnarray}
with the boundary conditions $h(0)=u(\infty) = 0$ and $u(0)= h(\infty)
=1$.  Substituting this Ansatz into the Euler-Lagrange equations of
the theory gives a set of coupled ordinary differential equations
that can be solved numerically.  Their solution is characterized by a
core region, of radius $R_{\rm core} \sim (ev)^{-1}$, outside of which
$u$ and $h$ approach their asymptotic values exponentially rapidly.
The total energy is 
\begin{equation}
    M_{\rm mon} \sim {Q_{\rm M}^2 \over R_{\rm core} } \sim {v \over e}   \, .
\end{equation}

We will find it useful to view this solution from a somewhat different
viewpoint \cite{Lee:1994sk}. To this end, note that by a singular
gauge transformation the fields of Eq.~(\ref{monoAnsatz}) can be
brought into the unitary gauge form
\eject
\begin{eqnarray}
      \varphi &=& h(r)   \cr\cr
      W_i &=& {f_i(\theta,\phi)\over er}\, u(r) \cr\cr
      {\cal A}_i  &=& {\cal A}_i^{\rm Dirac}   
\label{stringansatz}
\end{eqnarray}
where the $f_i(\theta,\phi)$ are complex functions whose explicit form
will not be needed and ${\cal A}_i^{\rm Dirac}$ is the U(1) Dirac
vector potential for a monopole of charge $1/e$.  (Because it is only
a gauge artifact, the string singularity of ${\cal A}_i^{\rm Dirac}$
will be of no concern.)  Note that $u(r)$ is directly related to the
magnitude of the charged vector meson field.

In this gauge, the structure of the monopole can be understood by
making reference to the form of the Lagrangian given in
Eq.~(\ref{unitaryLag}).  Thus, we can imagine constructing the
monopole solution by beginning with a point Dirac monopole.  Because
of the $1/r^2$ Coulomb magnetic field, this has a divergent energy
density near the origin.  However, this divergence can be cancelled by
introducing the charged vector field, provided that the magnetic
moment of the latter is properly oriented.  Indeed, the appearance of
a nonzero $W$ field is to be expected whenever the energy gain from
the interaction of the magnetic moment with the magnetic field
outweighs the cost in mass energy; in the presence of our Coulomb
field, this is the case for $r \lesssim (ev)^{-1} \sim R_{\rm core}$.
Finally, the vanishing of $\phi$ at the center of the monopole, which
is explained on topological grounds in the usual description, occurs
here because it minimizes the contribution of the $W$ mass term to the
energy.

The lesson to be drawn from this is that the appearance of a nonzero
$W$ field can be understood in terms of ``local'' physics, without any
reference to the topological behavior at spatial infinity.  In other
words, the value of $W_i({\bf r})$ at a given point is directly
related to the value of the magnetic field at that point.

\section{Self-gravitating monopoles and magnetically charged black
holes with hair} 
\label{sec:gravmono}

Let us now include gravity in this analysis.  One indication of what
to expect can be gained by noting that the Schwarzschild radius
$2MG \sim v/(eM_{\rm Pl}^2)$ is comparable to the core radius if
$v \sim M_{\rm Pl}$.  Hence, we might expect the monopole solutions to
become black holes when $v$ is greater than some critical value of the
order of the Planck mass.  (As long as $e \ll 1$, the mass and
horizon radius will be much greater than the Planck mass and Planck
length, respectively, so that quantum gravity
effects should be negligible.)   We will also see that these
monopoles can have related black hole solutions even when $v \ll M_{\rm
Pl}$.  

Let us begin by adapting the Ansatz of
Eq.~(\ref{monoAnsatz}) to a curved spacetime with a spherically symmetric
metric of the form of Eq.~(\ref{sphsymmetric}).  The matter field part of
the action can then be written in the ($1+1$)-dimensional form
\begin{equation}
    S_{\rm matter} = - 4\pi \int dt \, dr \, r^2 \sqrt{AB} \left[{K\over 
    A} + U \right] 
\label{reducedaction}
\end{equation}
where 
\begin{equation}
    K = {(u')^2 \over e^2 r^2} + {1\over 2} v^2(h')^2 
\label{Kdef}
\end{equation}
and 
\begin{equation}
    U = {(1-u^2)^2 \over 2 e^2 r^4} + {u^2 h^2 v^2 \over r^2} 
   + {\lambda v^4 \over 2}(1-h^2)^2   \, .
\end{equation}
One can view $U$ as being an $r$-dependent potential for two scalar
fields $u$ and $h$.  At large $r$, its minimum occurs when $u=0$ and
$h=1$.  Near the origin, it is minimized by $u=1$ and $h=0$.  For
small scalar self-coupling, $\lambda < e^2$, these are the only minima
of $U$.  However, if $\lambda > e^2$ there is an intermediate region
of $r$ where the potential has a nontrivial $r$-dependent minimum that
I will denote by $\hat u(r)$ and $\hat h(r)$.

The matter field equations can be obtained by varying 
the reduced action of Eq.~(\ref{reducedaction}).  This gives
\begin{equation}
    {1 \over \sqrt{AB} }\left({\sqrt{AB} u' \over A} \right)' 
         = {e^2 r^2 \over 2} {\partial U\over \partial u}
\label{ueq}
\end{equation} 
\begin{equation}
    {1 \over r^2\sqrt{AB}} \left({r^2\sqrt{AB} h' \over A} \right)' 
         = {1 \over v^2} {\partial U\over \partial h}   \, .
\label{heq}
\end{equation} 
These must be supplemented by equations for the metric functions $A$ and 
$B$.  Einstein's equations reduce to 
\begin{equation}
    {\cal M}' = 4 \pi r^2 \left( {K \over A} + U \right)
\label{Meq}
\end{equation} 
\begin{equation}
      { (AB)' \over AB} = 16 \pi G r K
\label{ABeq}
\end{equation}
where the mass function ${\cal M}(r)$ is defined by 
\begin{equation}
    {1 \over A(r)} = 1 -{ 2 G {\cal M}(r) \over r}   \, .
\end{equation} 

By substituting Eq.~(\ref{ABeq}) into Eqs.~(\ref{ueq}) and
(\ref{heq}), we can eliminate $B(r)$ and obtain a set of three coupled
differential equations for $u$, $h$, and $A$. These are subject to a
number of boundary conditions.  At spatial infinity, finiteness of the
energy requires that $u(\infty)=0$ and $h(\infty)=1$.  In order that the 
fields and metric be nonsingular at the origin, we must require that
$u(0)=1$ and $h(0)={\cal M}(0)=0$.  Finally, the coefficients of $u''$
and $h''$ in Eqs.~(\ref{ueq}) and (\ref{heq}) vanish at any zeroes of
$1/A$.  As a result, these equations give two additional
constraints among $u$, $h$, $u'$, and $h'$ at every horizon.

In general, a set of one first-order and two second-order equations
will allow at most five boundary conditions to be satisfied.  Hence,
we might hope to find solutions without horizons that are regular at
both the origin and infinity (i.e., nonsingular self-gravitating
monopoles) or black hole solutions that are finite at spatial infinity
and at one horizon, but singular at the origin.  Only for special
choices of parameters would we expect to be able to have solutions
that are regular at two horizons (like the Reissner-Nordstrom metric)
or solutions regular at a horizon and at both $r=0$ and $r=\infty$.

Of course, the presence of the correct number of boundary conditions
does not guarantee the existence of a solution.  To see whether there
actually is a solution, one must resort to numerical techniques
\cite{Lee:1992vy,Breitenlohner:1992aa,Breitenlohner:1995di,Ortiz:1992eu}.
One finds that $1/A$ develops a minimum at a value of $r$ of order
$v/e$.  As $v$ is increased this minimum becomes deeper until, at a
critical value $v_{\rm cr}$ of order $M_{\rm Pl}$, an extremal horizon
appears; this critical value varies with $\lambda/e^2$.  For $v >
v_{\rm cr}$ there are no nonsingular solutions.  Later in these
lectures I will return to these critical solutions and discuss the
approach to the black hole limit in more detail.

There are also solutions with horizons.  One type is obtained
trivially.  Setting $u=0$ and $h=1$ everywhere clearly satisfies
Eqs.~(\ref{ueq}) and (\ref{heq}).  Equations~(\ref{Meq}) and
(\ref{ABeq}) then lead to a Reissner-Nordstrom metric with magnetic
charge $1/e$ and arbitrary mass $M$.

We can also look for solutions with a horizon, but with nontrivial
matter fields outside the horizon; i.e., black holes with hair.  One
can imagine doing this by putting a small Schwarzschild-like black
hole in the center of a monopole.  In other words, we assume that
there is a horizon at $r_{\rm H} = 2G{\cal M}_0$, where ${\cal M}_0
\ll M_{\rm mon} \sim v/e$.  Because such a horizon would correspond to
a very light black hole, one would expect that its gravitational
effects outside the horizon would be small, and that for $r \gtrsim
r_{\rm H}$ the solution would be similar to that for the nonsingular
monopole.  This expectation is borne out by detailed numerical and
analytic investigations
\cite{Lee:1992vy,Breitenlohner:1992aa,Breitenlohner:1995di}.

These solutions are possible only in a certain region of
parameter space.  Thus, consider integrating Eq.~(\ref{Meq}) to obtain
the monotonically increasing mass function
\begin{equation} 
   {\cal M}(r) = {\cal M}_0 
    + 4\pi \int_{r_{\rm H}}^r ds \, s^2 \left( {K \over A} + U \right)   \, .
\end{equation} 
There will be a horizon whenever ${\cal M}(r)/r = 1/(2G)$.  By
construction, this occurs at $r=r_{\rm H}$.  If $v$ is small, ${\cal
M}(r)/r$ will initially decrease with increasing $r$ outside the
horizon, 
but will then begin to increase and, when $r \sim R_{\rm
core} \sim 1/(ev)$, reach a maximum of order $M_{\rm mon}/R_{\rm core} \sim
v^2$, after which it decreases and asymptotically vanishes. 
As $v$ is increased, the height of the maximum of ${\cal M}(r)/r$ will
increase until it reaches $1/2G$ at a critical $v$ of order $M_{\rm Pl}$. 
Since we do not expect to be able to find solutions regular at two
horizons and infinity, this sets a maximum value of $v$ for the
given ${\cal M}_0$.

This analysis assumes that $r_{\rm H}$ is well inside the monopole
core; we would not expect to find solutions with hair if $2G{\cal
M}_0$ were considerably larger than $R_{\rm core}$.  This leads to the
additional constraint ${\cal M}_0 \lesssim M^2_{\rm Pl}/(ev) \sim
M^2_{\rm Pl}/(e^2 M_{\rm mon})$.  More detailed
discussions and numerical analyses of these bounds are given in 
\cite{Lee:1992vy,Breitenlohner:1992aa,Breitenlohner:1995di}.

It is easy to see that these rough bounds allow the existence of
solutions with hair that have masses greater than the extremal
Reissner-Nordstrom mass $\sqrt{4 \pi} M_{\rm Pl}/e$.  This implies
that there can be two distinct black hole solutions with the same mass
and charge: the solution with hair, and the  
Reissner-Nordstrom solution.  This disproves the weaker form of the
no-hair conjecture.  It also raises the possibility of a transition
from one solution to the other.

\section{Instability of the Reissner-Nordstrom solution}
\label{sec:instability}

To explore this possibility, let us examine the stability under small
perturbations of a Reissner-Nordstrom solution with magnetic charge
$1/e$ and outer horizon radius $r_{\rm H}$ \cite{Lee:1992qs}.  For the
moment I will consider only spherically symmetric modes and write
\begin{eqnarray}
     A &=& A_{\rm RN}(r) + \delta A(r,t)   \cr 
     B &=& B_{\rm RN}(r) + \delta B(r,t)   \cr 
     h &=& 1 + \delta h(r,t)   \cr
     u &=& u(r,t) \,\, .
\end{eqnarray}
Linearizing the field equations in the perturbations $\delta A$,
$\delta B$, $\delta h$, and $u$, we find that they separate into a
pair of equations involving only $\delta A$ and $\delta B$, another
involving only $\delta h$, and one involving $u$.  It is clear that
the first set give no instability, since otherwise the
Reissner-Nordstrom solution would be unstable in the Maxwell-Einstein
theory, which we know is not the case.  It is also easy to see that
$\delta h$ has no unstable modes.  Hence, we need only consider $u$,
which obeys
\begin{equation}
    {1 \over \sqrt{AB}} {\partial \over \partial t} \left( {\sqrt{AB} 
    \over B} {\partial u\over \partial t}\right)
   - {1 \over \sqrt{AB}} {\partial \over \partial r} \left( {\sqrt{AB} 
    \over A} {\partial u\over \partial r}\right) = {u(1-u^2) \over r^2} 
    -e^2h^2v^2 u  \, \, .
\end{equation} 
Using the properties of the unperturbed metric and keeping only terms
linear in $u$, we obtain from this
\begin{equation}
    {1 \over B_{\rm RN}} {\partial^2 u\over \partial t^2}
   -{\partial \over \partial r} \left(B_{\rm RN} {\partial u\over \partial 
   r}\right) = - \left( e^2 v^2 - {1 \over r^2} \right) u    \, .
\label{instabeq}
\end{equation}   
An instability would correspond to an exponentially growing solution; i.e.,
a solution of the form $u = f(r) e^{i \omega t}$ with imaginary frequency
$\omega$.
 
The equation can be recast in a more familiar form by defining a new
coordinate $x$ by
\begin{equation}
      {dr \over dx} = B_{\rm RN}(r)   \, .
\label{xFROMr}
\end{equation}
This maps the exterior region, $r_{\rm H} < r <\infty$, onto the entire
real line, $-\infty < x < \infty$, and allows us to rewrite 
Eq.~(\ref{instabeq}) in
the form
\begin{equation}
    -{d^2 u \over dx^2} + V(x) u = - {d^2u \over dt^2} = \omega^2 u
\label{schroed}
\end{equation}  
where
\begin{equation}
    V(x) = {B_{\rm RN}(r) \over r^2} ( e^2 v^2 r^2 -1)
\end{equation}
with $r$ given as a function of $x$ through Eq.~(\ref{xFROMr}).  The
precise shape of $V$ depends on the value of $r_{\rm H}$, but in all cases
$V(-\infty) = 0$ and $V(\infty) = e^2v^2$.

Equation~(\ref{schroed}) is of the form of a non-relativistic
Schroedinger equation, and the existence of an instability is
equivalent to having a negative energy bound state.  This is
determined by the value of $r_{\rm H}$.  If $r_{\rm H} > 1/(ev)$, then $V(x)$ is
everywhere positive and there are no bound states.  If instead $r_{\rm H} <
1/(ev)$, there is a range of $x$ where $V(x)$ is negative and a bound
state becomes a possibility; numerical analysis shows that this
actually happens if 
\begin{equation}
     r_{\rm H} < {c \over ev}
\label{inequality}
\end{equation}
with $c \approx 0.557$ \cite{Ridgway:1995sm}.  While linear analysis
cannot by itself determine the eventual outcome of the instability, it
seems clear that the result is a static magnetically charged black
hole with $W$-boson hair.

Note that this stability analysis did not make use of the full
structure of the Yang-Mills theory, but only relied on the existence
of a charged vector field with a magnetic moment.  Recalling the
nontopological analysis of the 't Hooft-Polyakov monopole outlined in
Sec.~\ref{sec:monopole}, we see that the physical origin of the
instability lies in the fact that in a sufficiently strong magnetic
field it is energetically favorable to produce a nonzero $W$ field.
Making the horizon radius small enough guarantees that there will be a
magnetic field of critical strength outside the horizon.

This instability has dramatic consequences for the ultimate fate of a 
magnetically charged black hole.  Because of quantum mechanical effects, 
black holes emit thermal radiation at a Hawking temperature
\begin{equation}
    T_{\rm H} = {\hbar \over 4 \pi} \left({ B' \over \sqrt{AB}}
    \right)_{r=r_{\rm H}} \, .
\end{equation}
For a Schwarzschild black hole this temperature is inversely
proportional to the mass, so that as the black hole radiates
its temperature increases without limit, leading
to complete evaporation in a finite time.  A Reissner-Nordstrom black
hole initially follows the same scenario.  However, as the
mass approaches the extremal value the temperature begins to
decrease, with $T=0$ in the extremal limit.  Hence, unless
the black hole loses its charge (e.g., by preferential Hawking
radiation of particles of one charge over the other), the radiation
will eventually cease and the black hole horizon will remain forever.

The instability we have found changes this scenario.  Initially, the
black hole radiates and loses mass, just as before.  This causes the
horizon to contract until the inequality (\ref{inequality}) is
satisfied. At this point, nonzero vector meson fields begin to appear
outside the horizon, producing a black hole with hair whose Hawking 
temperature,
like that of a Schwarzschild black hole, never vanishes.  Radiation
continues unimpeded until the horizon has contracted to a point,
leaving behind a nonsingular monopole \cite{Tamaki:2000rf}.

\section{Static black holes without spherical symmetry}

One of the most striking results in classical black hole physics is
the fact that in the Einstein-Maxwell theory all static black holes
are spherically symmetric \cite{Israel-Sch,Israel-RN}.  In contrast
with electrodynamics, where static solutions corresponding to point
multipoles of arbitrary order are possible, higher ``mass multipoles''
seem to be ruled out.  By extending the analysis of the previous
section to Reissner-Nordstrom solutions with $q>1$ units of magnetic
charge, we can show that this is not always the case.

It is most convenient to work with the unitary gauge fields $\varphi$,
$\cal A_\mu$, and $W_\mu$.  The unperturbed solution has $\varphi({\bf
r}) = v$ and $W_\mu({\bf r}) = 0$ everywhere, while the metric and
electromagnetic field are those of a pure Reissner-Nordstrom solution
with magnetic charge $q/e$. It is natural to expand the perturbations
in spherical harmonics of appropriate types.  For the scalar field,
the electromagnetic field, and the metric, these are the standard
scalar, vector, and tensor spherical harmonics.  However, the
expansion of the charged vector field must be modified.  Recall that
in the presence of a magnetic charge $Q_{\rm M} = q/e$ a particle
carrying electric charge $e$ acquires an additional angular momentum
of magnitude $eQ_{\rm M} = q$ directed along the line from the
particle to the magnetic charge.  Because this is perpendicular to the
ordinary orbital angular momentum ${\bf r} \times m{\bf v}$, the
angular momentum of a spinless particle has a lower bound ${\bf L}^2
\ge q^2$.  Correspondingly, in the expansion of a charged scalar field
the usual spherical harmonics $Y_{LM}(\theta, \phi)$ must be replaced
by monopole spherical harmonics \cite{Tamm,Wu:1976ge} ${\cal
Y}_{q LM}$, with $L = q, q +1, \dots$ and $M= -L, -L+1, \dots, L$.
For a charged vector field (or more precisely, for its spatial
components) one must introduce monopole vector harmonics labeled by a
total angular momentum $J$.  Since this is the result of adding unit
spin angular momentum to the orbital angular momentum $L$, we can have
$J = L-1$, $L$, or $L+1$.  We therefore obtain vector monopole
spherical harmonics \cite{Weinberg:1994sg,Olsen:1990jm} ${\bf
C}^{(\lambda)}_{qJM}$, where $J = q-1, q, q +1, \dots$ and $\lambda$
distinguishes between different harmonics with the same values of $J$
and $M$.  There are three such harmonics for most values of $J$, but
for $J=q-1$ there is only a single multiplet of vector harmonics.

Note that $J=0$ can occur only if $q=1$, so that a spherically
symmetric $W$ field is possible only for unit magnetic charge.  This
explains the old result \cite{Weinberg:1976eq} that no finite energy SU(2)
configuration with multiple magnetic charge can be spherically
symmetric.  It also implies that any instability of the higher-charged
Reissner-Nordstrom solutions must lead to a solution with
non-spherically symmetric hair.

We now substitute the spherical harmonic expansions of the 
various fields into the action, keeping only terms quadratic in the 
perturbation.  Because the unperturbed solution is spherically symmetric, 
the quadratic action splits into a sum of terms with different angular 
momentum.   Each of these, in turn, splits into a part containing the 
metric and electromagnetic field perturbations, a part containing the 
scalar field perturbations, and a part involving only the perturbations of 
the massive vector field.   As was noted previously, 
we know that the first of these cannot give any
instability.  It is easy to see that the second term is also positive 
definite.  Thus, as with the singly-charged case of 
Sec.~\ref{sec:instability}, 
the only possible instability arises from the massive vector modes.  

Once again, the presence of an instability is equivalent to the existence
of a bound state in a Schroedinger-like problem.  However, the analysis is
more complicated than previously because there is more than one mode with
the same values of $J$ and $M$ for $J \ge q$.  Nevertheless, one still
finds \cite{Ridgway:1995sm}
that for all values of $J$ there is a bound state, and thus an unstable
mode, if the horizon radius $r_{\rm H}$ is less than a critical value $r_{\rm
cr}(J)$.  The largest $r_{\rm cr}(J)$ occurs for the minimum angular
momentum, $J=q-1$.  Hence, a Reissner-Nordstrom solution with horizon radius 
just
less than $r_{\rm cr}(q-1)$ has a single multiplet of $2q-1$ normalized
negative eigenmodes $\delta W_\mu = \psi_\mu^M$ that obey a differential
equation of the form
\begin{equation}
    {\cal M}_\mu{}^\nu  \psi_\nu^M = -\beta^2 m^2\psi_\mu^M  
\end{equation}
where $m$ is the unperturbed mass of the black hole and $\beta$ is 
dimensionless.  The solution is therefore classically unstable against
decay into a black hole with vector meson hair.  Because $J=q-1 \ne 0$, the
latter solution cannot be spherically symmetric.  It could, however, be
axially symmetric if, e.g., only the mode with $M=0$ were excited.  Other
combinations of modes, on the other hand, could lead to solutions with less
symmetry, or possibly no rotational symmetry at all.

To see which of these is the case, we must go beyond this linear
analysis \cite{Ridgway:1995ke}.  If we assume that $r_{\rm H}$ is just
below the critical value for instability, so that $\beta$ is small, we can
use a perturbative approach.  Let
\begin{equation}
     W_\mu = a\,m^{-1/2}\sum_M k_M \psi_\mu^M  + \tilde W_\mu 
      \equiv V_\mu + \tilde W_\mu
\label{Wpert}
\end{equation}
where $\tilde W_\mu$ is orthogonal to the negative modes.  The constants 
$k_M$ determine the angular dependence of the solution; they are assumed 
to be normalized so that
\begin{equation}
    \sum _{M=-q+1}^{q-1} |k_M|^2 = 1 \, .
\end{equation}
We will see that the overall scale $a$ is proportional to $\beta/e$.

Substituting Eq.~(\ref{Wpert}) into the $W$-field equations, one finds
that $\tilde W_\mu$ is of order $e^2 a^3$.  Maxwell's equations show
that the perturbation $\delta A_\mu$ of the electromagnetic field is
of order $e a^2$, while from Einstein's equations we find that the
metric perturbation $h_{\mu\nu} = O(G m^2 a^2)$.  If we assume that $G
m^2 \ll e^2$, the dominant terms in the matter Lagrangian can be
written schematically as
\begin{equation}
    {\cal L}_{\rm matter} = - V^{\mu *} {\cal M}_{\mu \nu} V^\nu  + 
     - e^2 V^4  + e(\delta{\cal A})V^2 + (\delta {\cal F})^2 
        +\cdots   \, .
\label{actionpert}
\end{equation}
The first term is of order $\beta^2 a^2$, the next three are $O(e^2a^4)$, 
and the omitted terms are suppressed by powers of either $a$ or $Gm^2/e^2$.

We now integrate Eq.~(\ref{actionpert}) over the region outside the
horizon.  Extremizing the resulting action with respect to $a$ shows
that $a$ is of order $\beta/e$.  By choosing $r_{\rm H}$ to be
sufficiently close to the critical value, we can make $a$ small enough
that the omitted terms in Eq.~(\ref{actionpert}) are indeed
negligible.  We must also minimize with respect to the $k_M$.  For the
$q=2$ doubly-charged black hole, this gives an axially symmetric
configuration.  This axial symmetry does not survive for larger $q$.
The $q=3$ and $q=4$ solutions have tetrahedral and cubic symmetries,
respectively.  (The somewhat surprising connection between these
regular polyhedra and magnetic charges can be understood in terms of
the number of zeroes of the $J=q-1$ vector harmonics
\cite{Ridgway:1995ke}.  Similar behavior is also found in other
contexts \cite{Braaten:1990rg,Houghton:1996bs}.)  For larger $q$,
there is in general no rotational symmetry at all.

At this point one can go back to the gravitational field equations
and determine the metric perturbations.  The angular dependence of the
matter fields gives rise to higher gravitational multipole moments,
with the consequent multipole fields only falling as powers of the
distance from the black hole.  However, despite the angular
dependence, the surface gravity remains constant on the horizon, just
as required by the zeroth law of black hole dynamics.

\section{The monopole-black hole transition}

Let us now return to the case of unit magnetic charge and examine in
more detail the transition from a nonsingular monopole to a
magnetically-charged solution with a horizon \cite{Lue:1999zp}.  In
this section I will focus on the extremal solutions that form the
boundary between these regimes, while in the next I will discuss the
solutions that are just short of this critical limit.

Because both $1/A$ and $(1/A)'$ vanish at the extremal horizon
$r=r_*$, it is a singular point of Eqs.~(\ref{ueq}-\ref{Meq}), and we
can expect to find nonanalytic behavior there.  Indeed, since $r$
itself is a singular coordinate at the horizon, in the sense that
there is an infinite metric distance from $r=r_*$ to any other value
of $r$, it would not be surprising if the derivatives of fields with
respect to $r$ were to diverge at the horizon.  Ordinarily, physical
considerations would determine the allowable singularities.  However,
here I am not actually requiring that the extremal solution be
physically acceptable, but only that it be the limit of a family of
physically acceptable nonsingular solutions.  With this in mind, I
will allow $u'$ and $h'$ to diverge, and will only require that this
divergence be such that $u'/\sqrt{A}$ and $h'/\sqrt{A}$ remain finite.
I will also assume that the leading singularities of the matter fields
and of the metric functions are of the form $|r-r_*|^\alpha$, with the
exponent $\alpha$ possibly being different on the two sides of the
horizon.

With these assumptions, Eqs.~(\ref{ueq}-\ref{Meq}) imply that at $r=r_*$ the 
matter fields must be at a stationary point of the 
$r$-dependent potential $U$, and that 
\begin{equation}
       1 - 8\pi G r_*^2 U(r_*) =0  \, .
\end{equation}
This allows two possible scenarios: In one, the horizon occurs at the
extremal Reissner-Nordstrom value $r_0 = \sqrt{4\pi G/e^2}$, and the matter
fields at the horizon are $u_*=0$ and $h_*=1$.  Since these values are
those expected far from the monopole core, where only the Coulomb fields
survive, I will refer to this case as having a ``Coulomb region horizon''. 
In the other possibility, $r_* < r_0$ and the matter fields have
nontrivial values $u_* = \hat u(r_*)$ and $h_* = \hat h(r_*)$ at the
horizon.  This gives an extremal solution with hair that I will refer to as
having a ``core region horizon''.

It was argued in Sec.~\ref{sec:gravmono} that there were too many boundary
conditions for one to expect a solution to be nonsingular at $r=0$,
$r=\infty$, and also at a horizon.  The nonanalyticity at an extremal
horizon invalidates this argument.  One effectively has two independent
boundary value problems to solve.  Integrating out from the horizon to
infinity, there are two free constants in the Taylor expansions of $u$ and
$h$ at the horizon that can be adjusted so as to satisfy the two boundary
conditions at $r=\infty$.  Integrating inward, one must be able to satisfy
three conditions at $r=0$.  The Taylor expansions of the matter fields at
$r_*$ (which are independent of the expansions on the other side of the
horizon) only provide two adjustable constants.  To obtain a third, we
recall that an extremal solution only arises when $v$ is at a critical
value $v_{\rm cr}$; hence, we can think of $v$ as being the third
adjustable constant.

Carrying out this analysis in detail, one finds two rather different 
behaviors.  With a Coulomb region horizon, the exterior solution is a pure 
Reissner-Nordstrom solution with $u=0$ and $h=1$.  Just inside the 
horizon, one finds that 
\begin{eqnarray}
     u &=& p|x|^{1/2}  + C_u |x|^{\gamma_u} + a x +\cdots  \cr\cr
     h &=& 1 - C_h |x|^{\gamma_h} + b x + \cdots   \cr\cr
     {1\over A} & =&  k x^2 + \cdots
\end{eqnarray}
where $x \equiv (r -r_*)/r_*$.  Here $p$, $k$, $a$, and $b$
are determined in terms of the parameters of the theory, as are the 
exponents $\gamma_u$ and $\gamma_h$, both of which are greater than $1/2$.  
The constants $C_u$ and
$C_h$ can be adjusted so that the boundary conditions at the origin are
satisfied.  The terms indicated by ellipses are determined by the terms 
shown explicitly.
Note that $k \ne 1$, so that $(1/A)''$ is
discontinuous at the horizon.  (Solutions with the square root singularity in 
$h$ rather than $u$ are also possible.)

The solutions with core region horizons behave more smoothly and do not 
have a square root singularity.  Near the 
horizon, 
\begin{eqnarray}
     u &=& \hat u(r_*) + a x + p_1 C_1 |x|^{\gamma_1} 
         + p_2 C_2 |x|^{\gamma_2} + \cdots \cr\cr
     h &=& \hat h(r_*) + b x + q_1 C_1 |x|^{\gamma_1} 
         + q_2 C_2 |x|^{\gamma_2} + \cdots   \cr\cr
     {1\over A} &=&  Fx^2 + \cdots \,\, .
\end{eqnarray}
The adjustable coefficients $C_1$ and $C_2$ can take on different values
inside and outside the horizon, while the other constants are fixed by the
parameters of the theory.  

Numerical integration of the field equations is needed to determine
which type of critical behavior actually happens in a particular case.
When $\lambda/e^2 \lesssim 25$, the critical solution has a Coulomb
region horizon \cite{Lue:1999zp,Brihaye:2000kt}.  For larger values of
$\lambda/e^2$, a core region horizon is found.  The approach to the
critical solution is rather curious in this case.  Initially, there is
a minimum in $1/A$ at $r \approx r_0$ that gets deeper as $v$
approaches $v_{\rm cr}$, just as if a Coulomb region horizon were
about to be formed.  However, just before $v_{\rm cr}$ is reached a
second minimum appears at a smaller value of $r$; it is this latter
minimum that becomes the extremal horizon.

To see what is perhaps the most striking difference between the two cases, 
we must return to Eq.~(\ref{ABeq}), which we have thus far ignored.  
Integrating this equation, we find that 
\begin{equation}
      (AB)_r = (AB)_\infty \exp\left[-16 \pi G \int_r^\infty ds\, s K 
      \right]
\end{equation}
where $K$, defined in Eq.~(\ref{Kdef}), contains the gradient terms in the 
dimensionally reduced matter Lagrangian.  This does not lead to anything 
particularly unusual when there is a core region horizon.  For a Coulomb 
region horizon, on the other hand, the square root singularity in $u$ (or 
$h$) leads to a divergence in the integral 
on the right hand side of this equation.  This gives a step-function rise 
in $AB$, so that for any two points 
$r_1 < r_*$ and $r_2 > r_*$
\begin{equation}
      { (AB)_{r_1} \over (AB)_{r_2} } = 0  \, .
\label{ABstep}
\end{equation}

This behavior at a Coulomb region horizon leads to a phenomenon
identified recently by Horowitz and Ross
\cite{Horowitz:1997uc,Horowitz:1998ed}.  It is often said that,
because the horizon is only a coordinate singularity rather than a
true curvature singularity, a freely-falling observer should feel no
unusual effects at the time of crossing the horizon.  In fact, the
acceleration of a radially infalling observer near the horizon can
invalidate this conclusion.  This can be seen by relating the
curvature components in a boosted frame where the observer is
instantaneously at rest to the components in a ``static'' frame where
the metric is time-independent.  Using orthonormal components in both
cases, we have
\begin{eqnarray}
   R_{t'kt'k} &=& R_{tktk} + \sinh^2 \alpha (R_{tktk} + R_{rkrk})  \cr
   R_{r'kr'k} &=& R_{rkrk} + \sinh^2 \alpha (R_{tktk} + R_{rkrk})  \cr
   R_{t'kr'k} &=&  \sinh \alpha \cosh \alpha (R_{tktk} + R_{rkrk})  \cr
   R_{t'r't'r'} &=& R_{trtr}
\end{eqnarray}
where primes denote coordinates in the infalling frame, $k$ indicates either 
transverse angular coordinate, and $\alpha$ is the boost factor.  Since 
$\alpha$ can become large as the observer nears the horizon, it is possible 
for the curvature components in the infalling frame (i.e., the components 
actually ``felt'' by the observer) to be large even though all components in 
the static frame are small.  The fact that this does not happen in the 
the Schwarzschild and Reissner-Nordstrom metrics is a 
consequence of the fact that these metrics have the special property that 
$R_{tktk} +  R_{rkrk} =0$.  

In an arbitrary metric of the form of Eq.~(\ref{sphsymmetric}), 
an infalling particle 
with an energy to mass ratio $E$ feels a tidal force proportional to
\begin{equation}
    R_{t'kt'k} = -{1 \over 2r} {d\over dr}\left[ {E^2 \over AB} 
     - {1 \over  A} \right]  \,\,.
\label{nakedBH}
\end{equation}
Horowitz and Ross exhibited several examples of dilaton black holes
for which this quantity
becomes large at exterior points near the horizon.  Because this
implies that an observer will feel a black hole-induced
``singularity'' even before crossing the horizon, they termed such
solutions ``naked black holes''.

Comparing Eqs.~(\ref{ABstep}) and (\ref{nakedBH}), we see that the
same phenomenon occurs for near-critical monopole solutions that are
close to developing a Coulomb region horizon.  As the extremal
solution is approached, the value of the right hand side of
Eq.~(\ref{nakedBH}) at the quasi-horizon diverges.  Hence, these
solutions become ``naked black holes'' even before they become black
holes.  It might be tempting to conclude that this singular behavior
is a necessary concomitant of the transition from a nonsingular
spacetime to one with a horizon.  However, the existence of solutions
with core region horizons at large $\lambda/e^2$ shows that this is
not the case.

\section{Quasi-black holes and the emergence of black hole entropy}

I now turn to solutions that are just short of being black holes;
i.e., solutions for which the minimum of $1/A$ has a value
$\epsilon$ that, while positive, is very close to zero.  These are
nonsingular and topologically trivial.  However, one might expect that
as $\epsilon$ decreases and the critical solution is approached, it
would be harder and harder for an external observer to distinguish
these solutions from true black holes.  Hence, it seems appropriate to
call these ``quasi-black holes'', and to denote the minimum of
$1/A$ at $r=r_*$ a ``quasi-horizon''.

Let us now consider how these solutions would appear to an observer
who remains at a radius $r \gg r_*$ \cite{Lue:2000qr}.  In order to determine
whether or not the solution was actually a black hole, the observer
could employ various means to try to probe the region inside the
quasi-horizon.  One possibility would be 
send in a particle and wait for it to emerge again.  Thus,
consider a massive particle moving on a geodesic that starts from an
initial radius $r_1 \gg r_*$ at time $t$, goes in to a minimum radius
$r_{\rm min} < r_*$, and then returns again to $r_1$ at a time
$t+\Delta t$.  Without loss of generality, we can assume that the
geodesic lies in the $\theta = \pi/2$ plane.  It will be characterized
by the conserved energy per unit mass $E= B(dt/d\tau)$ and angular
momentum per unit mass $J = r^2 (d\phi/d\tau)$.  A standard
calculation then gives
\begin{equation}
    {dr \over d\tau} = { 1\over \sqrt{AB}} 
    \left[ E^2 - B\left({J^2 \over r^2} + 1 \right) \right]^{1/2} \,\, .
\end{equation}
Integrating $dt/dr = (dt/d\tau)/ (dr/d\tau)$ gives the elapsed 
coordinate time
\begin{equation}
    \Delta t = 2 \int_{r_{\rm min}}^{r_1} dr \, {A \over \sqrt{AB}} 
   \left[ 1 - {B \over E^2}\left({J^2 \over r^2} + 1 \right) \right]^{-1/2}
    \,\, .
\end{equation}

For a solution with a core region quasi-horizon, the integral is dominated by 
the region $r \approx r_*$, and 
\begin{equation}
    \Delta t \approx k_1 { r_* \over \sqrt{(AB)_{r_*}}} \,\epsilon^{-1/2}
\end{equation}
where $k_1$ is a constant of order unity.

In the Coulomb horizon regime, the interior dominates the integral and 
\begin{equation}
     \Delta t \approx k_2 \, r_* \, \epsilon^{-q}
\end{equation}
where $k_2$ is also of order unity and $0.7 < q < 1$.  Curiously, although 
the coordinate time needed to traverse the interior diverges in the 
critical limit, for this case
the proper time vanishes as $\epsilon^q$.  

One could also probe the quasi-black hole by scattering waves off of it.  
Consider, for 
example, a massive scalar field $\phi$ obeying the curved space 
Klein-Gordon equation
\begin{equation}
   0 = { 1\over \sqrt{g}}\, \partial_\mu \left[ \sqrt{g}\, g^{\mu\nu} 
   \,\partial_\nu \phi\right]  + m^2 \phi   \, .
\end{equation}
By defining $\psi = r \phi$ and introducing a new coordinate $y$ obeying
\begin{equation}
   {dr \over dy} = {\sqrt{AB}\over A}
\end{equation}
we can rewrite this equation as
\begin{equation}
    0 = {\partial^2 \psi \over \partial t^2} 
        - {\partial^2 \psi \over \partial y^2} 
          + [U(r) + m^2 B] \psi
\end{equation}
where the potential
\begin{eqnarray}
     U &=& {1 \over 2r}{d \over dr} \left[{AB \over A^2} \right]
             + {J(J+1) \over r^2} B   \cr\cr
       &=& {AB \over r A} \left[ {8 \pi G K \over A} -{d \over dr} 
       \left({1 \over A}\right) + {J(J+1) \over r} \right] \,\, .
\end{eqnarray}

For either type of critical solution $U(r_*)$ tends to zero as $\epsilon 
\rightarrow 0$.  As a result, there is a clear distinction between the 
reflected wave arising from interaction with the potential in the region 
$r > r_*$ and that arising from interactions in the region $r < r_*$.  As 
the critical limit is approached, the former becomes indistinguishable 
from the wave reflected by the corresponding black hole solution.  The 
existence of a reflected wave from the interior region, as well as of a 
transmitted wave, distinguishes the nonsingular monopole from the black 
hole.  However, both of these suffer a time delay proportional to either 
$\epsilon^{-1/2}$ or $\epsilon^{-q}$, just as for the particle probe.   

An external observer with unlimited time available would be able to use 
probes such as these to gain information about the interior region of the
quasi-black hole.  However, any real observer must work on some finite time
scale $\Delta\cal T$.  For such an observer, the interior is inaccessible
if $\epsilon$ is too small [less than $(\Delta {\cal T})^{-2}$ or $(\Delta
{\cal T})^{-1/q}$ for the core- and Coulomb-type solutions, respectively]. 
The natural way to describe the system would then be by means of a density
matrix $\rho$ that was obtained by tracing over the degrees of freedom in
the interior.  This in turn would give rise to a naturally defined entropy
\begin{equation}
    S_{\rm QBH} = - {\rm Tr} \, \rho \ln \rho  
\end{equation}
for the quasi-black hole

An estimate of the value of this entropy can be obtained from a
calculation by Srednicki \cite{Srednicki:1993im}.  He considered a
free massless scalar field in a flat spacetime.  Assuming that the
system was in its ground state, by tracing over the degrees of freedom
inside an arbitrary spherical region he obtained an entropy
\begin{equation}
     S = \kappa M^2 A
\end{equation}
where $\kappa$ is a numerical constant of order unity, $M$ is an 
ultraviolet cutoff, and $A$ is the area of the boundary of the region.
In a gravitational context, we expect the Planck mass to provide the 
ultraviolet cutoff.  Hence, it is quite plausible that in the 
critical limit the 
entropy of the quasi-black hole will approach the 
Bekenstein-Hawking black hole entropy $(1/4) M_{\rm Pl}^2 A$.  However, in 
contrast with the black hole case, the quasi-black hole is topologically 
trivial.  Its ``interior'' is nonsingular and static.  Furthermore, this 
region is unambiguously defined, so that it is at least conceptually clear 
what it means to trace over the interior degrees of freedom.

\section{The third law of black hole dynamics}

The third law of thermodynamics has several formulations, one of which
states the impossibility of reaching zero temperature in a finite
time.  Since extremal black holes have zero Hawking temperature, the
analogies between thermodynamics and black hole dynamics then suggest
that they should be difficult, if not impossible, to create.  Indeed,
one formulation of the third law of black hole dynamics states that,
under certain technical assumptions, a nonextremal black hole cannot
be made extremal \cite{Israel-ThirdLaw}.  The essential difficulty can
be understood by recalling that an extremal Reissner-Nordstrom black
hole has a mass and a charge that are equal in Planck units, whereas a
nonextremal black hole has greater mass than charge.  If one tried to
make a nonextremal black hole extremal by dropping in matter with more
charge than mass, the Coulomb repulsion would tend to overcome the
gravitational attraction.
    
One could also try to produce an extremal black hole by starting with
a nonsingular spacetime and allowing the collapse of a shell of matter
with a properly adjusted mass to charge ratio.  Boulware showed that
this could in fact be done \cite{Boulware}.  However, this mechanism
relies crucially on the shell being infinitely thin; it fails for
shells of finite thickness.

The quasi-black holes of the previous section suggest another
possibility.  Because of the cancellation of the Coulomb energy in the
core by the magnetic dipole interaction, these monopoles have greater
charge than mass.  To bring them to criticality and produce an
extremal black hole, it should only be necessary to drop in an
appropriate amount of uncharged matter.  With a Coulomb region
quasi-horizon, one might run into difficulties from the naked black
hole behavior.  However, these can be avoided by working in the high
$\lambda/e^2$ regime where the critical solution has a core region
horizon.  A variation on this process starts with a solution
containing a small black hole in the center of the monopole core and
an almost critical quasi-horizon further out in the monopole core.
Here, the effect of the infalling matter is to replace the initial
finite temperature horizon by a zero temperature horizon at a larger
value of $r$.  This scenario has been tested by numerical simulations
using a massive neutral scalar field as the infalling matter
\cite{Lue:2000qr}.  The results of these are completely consistent
with expectations.  The possibility of such processes should give us
clues for a more precise formulation of the third law of black hole
dynamics.

\bigskip

This work was supported in part by the U.~S.~Department of Energy.

\end{document}